\documentclass[12pt,preprint]{aastex}

\usepackage{hyperref}
\usepackage[all]{hypcap}
\usepackage{graphicx}
\usepackage{mathrsfs}
\usepackage{natbib}
\usepackage{float}
\usepackage{amssymb}
\usepackage{color}
\usepackage{captcont}

\usepackage{amsmath}
\newcommand{\e}            {\mbox{$^{-1}$}}
\newcommand{\ee}           {\mbox{$^{-2}$}}

\newcommand{\kms}          {\mbox{${\rm km~s^{-1}}$}}
\def\twelveCO{\mbox{$^{12}$CO}}
\def\13CO{\mbox{$^{13}$CO}}
\def\C18O{\mbox{C$^{18}$O}}

\begin{document}

\title{Measuring protoplanetary disk gas surface density profiles with ALMA}
\author{Jonathan P. Williams and Conor McPartland}
\affil{Institute for Astronomy, University of Hawaii at Manoa, Honolulu, HI, 96822, USA; jpw@ifa.hawaii.edu}
\shorttitle{Gas Surface Densities of Protoplanetary Disks}
\shortauthors{Williams \& McPartland}

\begin{abstract}
The gas and dust are spatially segregated in protoplanetary disks
due to the vertical settling and radial drift of large grains.
A fuller accounting of the mass content and distribution in disks
therefore requires spectral line observations.
We extend the modeling approach presented in
\citet{2014ApJ...788...59W}
to show that gas surface density profiles can be measured
from high fidelity \13CO\ integrated intensity images.
We demonstrate the methodology by fitting
ALMA observations of the HD\,163296 disk to determine
a gas mass, $M_{\rm gas} = 0.048\,M_\odot$,
and accretion disk characteristic size
$R_c = 213$\,au and gradient $\gamma = 0.39$.
The same parameters match the \C18O\ 2--1 image and indicates an
abundance ratio [\13CO]/[\C18O] of 700 independent of radius.
To test how well this methodology can be applied to future line surveys
of smaller, lower mass T Tauri disks, we create a large \13CO\ 2--1 image
library and fit simulated data.
For disks with gas masses $3-10\,M_{\rm Jup}$ at 150\,pc,
ALMA observations with a resolution of $0\farcs 2 - 0\farcs 3$
and integration times of $\sim 20$ minutes allow reliable estimates
of $R_c$ to within about 10\,au and $\gamma$ to within about 0.2.
Economic gas imaging surveys are
therefore feasible and offer the opportunity to open up a new dimension
for studying disk structure and its evolution toward planet formation.
\end{abstract}

\keywords{circumstellar matter — planetary systems: protoplanetary disks
— solar system: formation}

\section{Introduction}
The gas and dust in circumstellar disks share a common origin in the
interstellar medium but rapidly evolve to a very different states.
The high densities, cool temperatures, and low turbulence in disks
provide the ideal conditions for the growth of dust grains to
millimeter sizes and beyond.
As the ratio of surface area to mass decreases, the grains feel a
headwind from the slightly sub-Keplerian motion of the viscous
gas and they drift inwards while also sedimenting toward the midplane.
A wealth of fascinating physical processes can then occur in this
two-fluid medium that are not found in other astronomical environments
and that typically end with a planetary system
\citep{2013apf..book.....A}.

Protoplanetary disks radiate most strongly in the infrared and measurements
of excess emission above the photosphere at these wavelengths is the most
sensitive way to diagnose the presence of dust around stars.
The emission at millimeter wavelengths is much weaker but the continuum is
generally optically thin and therefore provides a more precise way to
measure the amount and, through interferometry, the distribution of the dust
\citep{2011ARA&A..49...67W}.

Molecular gas, though by far the dominant constituent by mass,
is actually much harder to observe since the gas is too cool for
H$_2$ to emit significantly.
As with observations of molecular clouds and cores,
the gas is most readily detected through millimeter observations
of rotational transitions of CO and other trace molecules
that lie within a warm molecular layer
\citep{2002A&A...386..622A}.
The high optical depth of
the infrared continuum obscures line emission which is otherwise
seen in the inner regions of transition disks with dust-depleted holes
\citep{2008ApJ...684.1323P}.
The dust becomes transparent at millimeter wavelengths
and many rotational lines can in principle be detected.
The \twelveCO\ lines are optically thick, however, and most useful as a
diagnostic of the temperature and kinematics of the gas than of its mass
\citep{1993ApJ...402..280B}.

\citet[][hereafter Paper I]{2014ApJ...788...59W}
showed the utility of CO isotopologue observations for
measuring disk gas masses independently from that of the dust.
The intensity of these rarer species, with their lower optical depths,
depends primarily on the amount of the gas and secondarily on its
temperature and density.
Freeze-out in the cold midplane and photo-dissociation in the
upper disk atmosphere must also be taken into account but we found
that most of the gas in typical disks resides in the molecular
layer between these two regions.
We concluded that the combination of spatially and velocity
integrated \13CO\ and \C18O\ line luminosities constrains disk gas
masses to a factor of about 3--10,
a result that has recently been confirmed in a much larger
Atacama Large Millimeter/Submillimeter Array (ALMA)
survey of Lupus disks by \citet{2016arXiv160405719A}.
This moderate level of precision
is sufficient to show that the bulk gas-to-dust ratios in most
protoplanetary disks are different than the interstellar medium
value of 100, a key finding that may help explain why the abundant
super-Earths and Neptunes in exoplanet surveys avoided
runaway growth to Jupiters \citep{2014ApJ...789...69H}.

In this paper, we examine whether we can extend the modeling methodology
in Paper I and use CO isotopologue maps to determine the distribution
of the gas. Techniques to determine dust surface density profiles from
resolved continuum images are now well established
\citep[e.g.,][]{1997ApJ...489..917L, 2007ApJ...659..705A}.
As the dust and gas are spatially decoupled, however,
we cannot simply extrapolate this to the gas distribution
\citep{2009A&A...501..269P}.
ALMA will make resolved disk images of CO isotopologues routine and
there is a need to develop simple modeling tools that can quickly and
reliably derive basic disk gas properties in a uniform way to
allow comparative studies in large surveys.
Our focus here is on observations of \13CO\ as it is strong enough to survey
and is more dependent on the surface density of the gas than its temperature.
\S\ref{sec:HD163296} describes the modeling procedure, the creation
of an image library, and an interpolation routine that allows a
continuous sampling of parameter space necessary for error estimation.
We then fit observations of the HD\,163296 disk as a proof-of-concept.
In the following section, \S\ref{sec:TTdisk_grid},
we create a generic grid of models more suitable for lower
mass disks around lower mass, T Tauri stars.
By comparing simulated images to gaussian fits and the rest of
the image library, we find that that ALMA integrations of a
few to a few tens of minutes (depending on gas mass)
at $\sim 0\farcs 2 - 0\farcs 3$ resolution should reveal
the gas surface density profiles of typical disks in
nearby star-forming regions.
We summarize these results and discuss their implications in
\S\ref{sec:discussion}.

\section{The HD 163296 disk}
\label{sec:HD163296}
Protoplanetary disks are generally compact with weak line emission.
Consequently, there are only a few disks with well resolved,
high signal-to-noise \13CO\ maps in the literature.
The exceptional disk around the nearby (122\,pc)
Herbig Ae star HD\,163296 is, fortunately, both big and bright.
Moreover, early ALMA observations of \13CO\ and other lines are
publicly available through its Science Verification
program\footnote{https://almascience.nrao.edu/alma-data/science-verification}
and it has been modeled independently by two groups
\citep{2013ApJ...774...16R, 2013A&A...557A.133D}.
It is therefore an ideal source to demonstrate the feasibility of
our procedure.

\subsection{Description of the model}
We follow the methodology in Paper I but create a library of
resolved images rather than a table of total line luminosities.
The gas is assumed to be azimuthally symmetric,
in hydrostatic equilibrium and under Keplerian rotation.
Nominally, there are nine free parameters but we are able to
considerably reduce these based on previous studies.
In particular, we set the
stellar mass, $2.3\,M_\odot$, and inclination, $i=45^\circ$,
based on \citet{2011ApJ...740...84Q} and we use the gas temperature
structure, $T_{\rm gas}(R,Z)$, described in \citet{2013ApJ...774...16R}.
This leaves just three remaining parameters,
$M_{\rm gas}$, $R_c$, and $\gamma$,
that determine the accretion disk gas surface density distribution
\citep{1974MNRAS.168..603L},
\begin{equation}
\Sigma_{\rm gas}(R)=(2-\gamma)\,\frac{M_{\rm gas}}{2\pi R_c^2}
                    \left(\frac{R}{R_c}\right)^{-\gamma}\,
                    \exp\left[-\left(\frac{R}{R_c}\right)^{2-\gamma}\right].
\end{equation}

Having specified the density and temperature structure, we define the
warm molecular layer where CO is in the gas phase and can emit with
a lower boundary set by a freeze-out temperature of 20\,K and an
upper boundary set by dissociation at column densities
$N_{\rm H_2} > N_{\rm dissoc} = 1.3\times 10^{21}\,{\rm H}_2$\,cm\ee.
Within this region, we assume a constant CO abundance
$\rm{[\twelveCO]/[H_2]} = 1\times 10^{-4}$
and isotopologue ratio, [\twelveCO]/[\13CO]=70.

We then calculate the \13CO\ 2--1 line emission using the
radiative transfer code
RADMC-3D.\footnote{http://www.ita.uni-heidelberg.de/$\sim$dullemond/software/radmc-3d/}
The output is a spectral line datacube with a resolution of 5\,AU
and 0.1\,\kms. The ordered motion of a Keplerian disk means that
different regions of a disk generally have different
radial velocities. As a result, we found that it is not necessary
to make tomographic comparisons of channel maps to discriminate between
models and that velocity integrated (zero-moment) maps suffice.
This reduces the computational requirements
of memory, disk space, and speed considerably.
Finally, we compared models with LTE and NLTE excitation and found
no substantial difference ($<1$\,mJy\,\kms\,pixel\e).

\subsection{Image interpolation}
\label{sec:image_interpolation}
We created an image library by running 748 models over the range
of parameters shown in Table~\ref{tab:HD163296_grid}.
It is straightforward to determine the best fit model parameters
by minimizing the squared difference between the image library
to the data. This simple chi-squared analysis does not allow us to
determine parameter errors, however, as the model is non-linear
\citep{2010arXiv1012.3754A}.
A more statistically robust approach is to sample the parameter
space using a Markov Chain Monte Carlo (MCMC) technique.
The radiative transfer calculation takes several minutes to
run for each set of disk parameters and is too slow to carry out
the required $\gg 10^4$ model calculations directly, however,
and we therefore designed a simple routine to interpolate images
within the model grid.

\capstartfalse
\begin{deluxetable}{cccc}
\tabletypesize{\footnotesize}
\centering
\tablecolumns{3}
\tablewidth{0pt}
\tablecaption{Parameter range of the HD\,163296 grid \label{tab:HD163296_grid}}
\tablehead{
\colhead{Parameter} & \colhead{Range} & \colhead{Step} & \colhead{Units}
}
\startdata
$M_{\rm gas}$       & 4.0--5.5 &   0.5   & $10^{-2}\,M_\odot$ \\
$R_c$               & 160--320 &  10     & au                 \\
$\gamma$            & 0.0--1.0 &   0.1   &                    \\
inclination         & 45       & \nodata & $^\circ$           
\enddata
\end{deluxetable}

We wish to determine the image $I({\bf p})$ at a set of parameter
values ${\bf p} = \{p_i;\ i=0, 1, ..., N\}$. For each index $i$,
we bound the parameter by grid points, $g_i^0 \leq p_i \leq g_i^1$.
There are $2^N$ vertices of the $N$-dimensional cube defined
by these grid points, ${\bf v} = \{g_i^j;\ i=0, 1, ..., N\}$
over all combinations $j=0,1$.
We then linearly interpolate the image library,
\begin{equation}
I({\bf p}) = \sum_{\bf v} w({\bf v})I({\bf v}),
\end{equation}
where the weights at each vertex,
\begin{equation}
w({\bf v}) = \prod_i \left(1 - \frac{|p_i - v_i|}{g_i^1 - g_i^0}\right).
\end{equation}

This procedure is fast and can be readily modified to allow different
weighting schemes or to extend over a wider parameter space beyond
the bounding grid points. It sufficed for our purposes where we found
that interpolated images averaged within 3\% of a full radiative
transfer calculation.
This efficient method to calculate model images over a continuous
range of parameter values now permits an MCMC analysis.

\subsection{Parameter estimation}
\begin{figure*}[tb]
\centering
\includegraphics[width=6.5in]{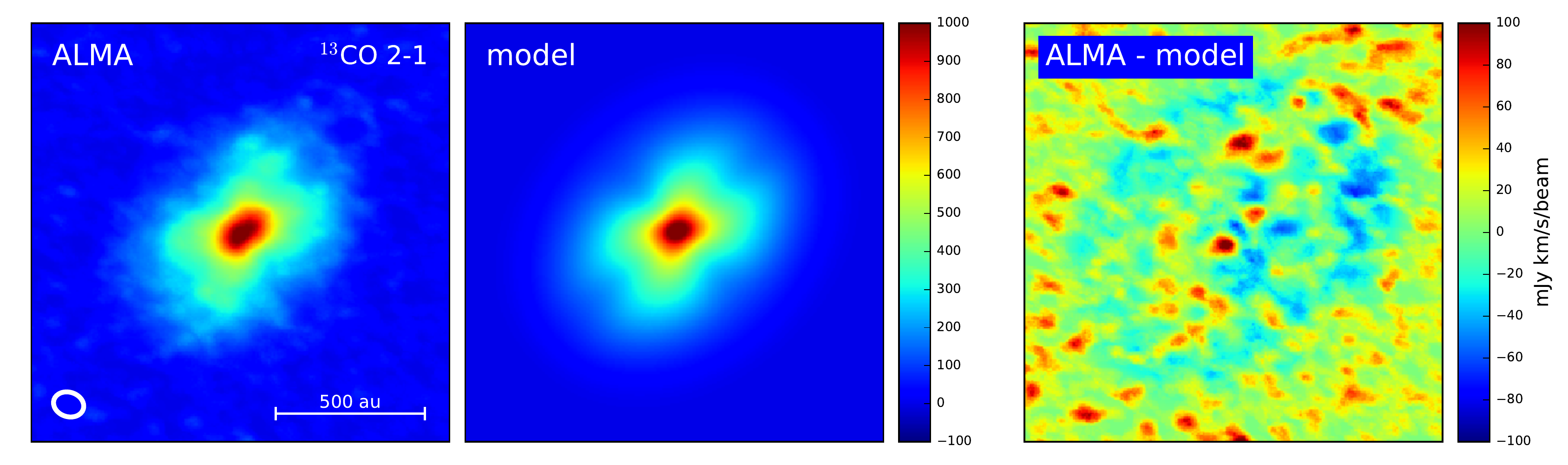}
\caption{
Comparison of \13CO\ integrated intensity images.
The left panel shows the ALMA Science Verification map,
the central panel the model fit for the median of MCMC
model parameters, and the right panel shows the difference image.
The colorbars show the range of intensities for each map;
the ALMA and model images on the same scale from
$-100$ to 1000\,mJy\,\kms\,beam\e,
and the difference image from $-100$ to 100\,mJy\,\kms\,beam\e.
}
\label{fig:MCMC_cornerplot}
\end{figure*}

The MCMC modeling was run with a flat prior for each parameter
over the range shown in Table~\ref{tab:HD163296_grid}
using the {\tt emcee} software package \citep{2013PASP..125..306F}.
The projection of the 3-dimensional parameter space is shown in
Figure~\ref{fig:MCMC_cornerplot}, from which we calculate median
and 68\% ($\pm 1\sigma$) confidence intervals, 
$M_{\rm gas}=0.048^{+0.003}_{-0.003}\,M_\odot,
 R_c=213^{+7}_{-7}\,{\rm AU},
 \gamma=0.39^{+0.09}_{-0.08}$.
The inferred gas mass agrees well with that derived from the
comparison of \13CO -- \C18O\ luminosities in Paper I ($0.047\,M_\odot$).
The apparently high precision obtained here reflects the rigidity of
the temperature structure imposed in our models, which we discuss
further in \S\ref{sec:discussion}.

\begin{figure}[tb]
\centering
\includegraphics[width=6.5in]{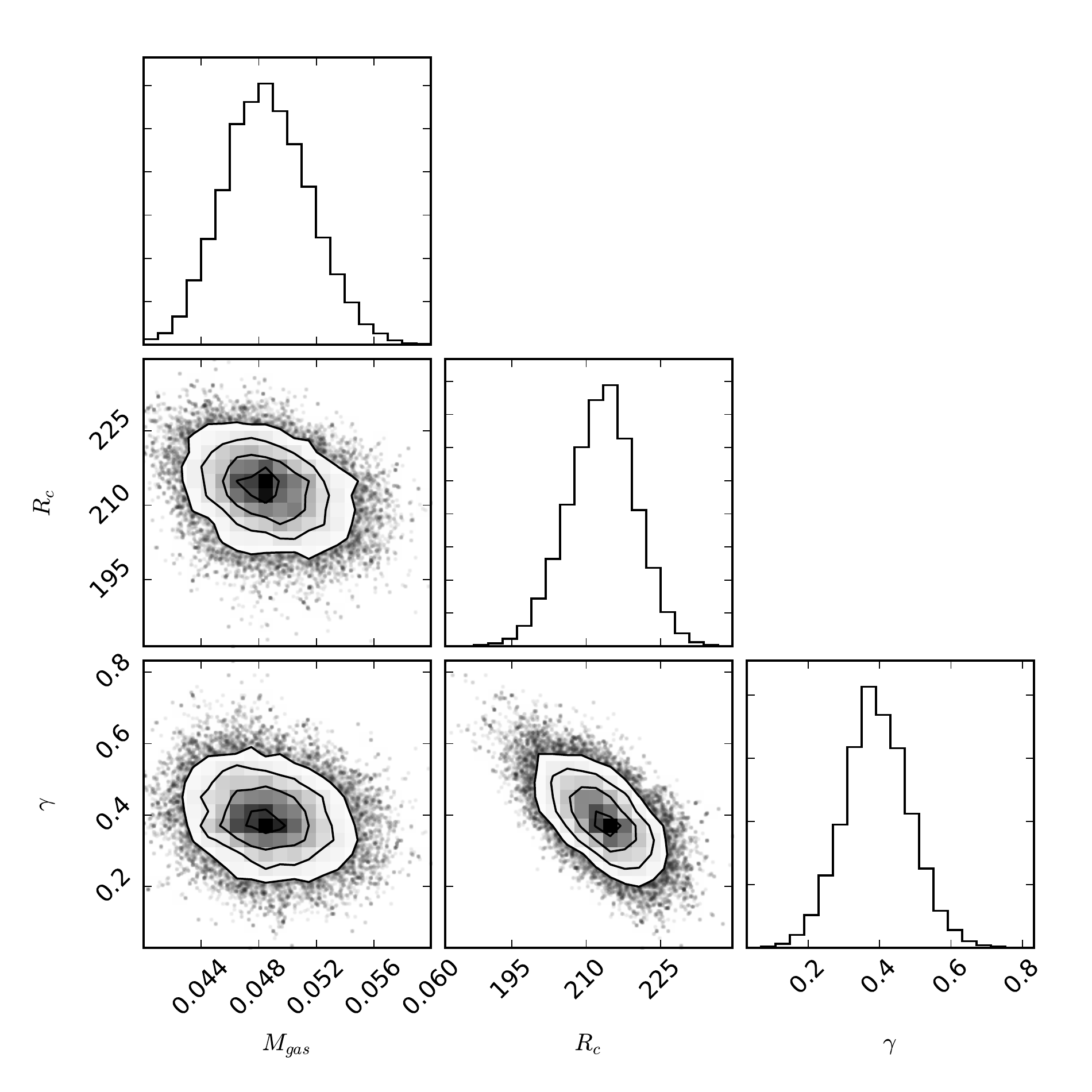}
\caption{
The distribution of posterior probabilities in the MCMC fitting
of the \13CO\ integrated intensity map of HD\,163296.
The three free parameters, gas mass, characteristic radius
and surface density gradient are all well constrained.
}
\label{fig:difference_image13}
\end{figure}

The data, median fit, and difference image
are shown in Figure~\ref{fig:difference_image13}.
Given the simplicity of the model, the overall fit is good with
peak residuals less than 10\% of the image maximum
and consistent with a gaussian with standard deviation 27\,mJy\,\kms,
comparable to the rms noise level of the data.
Due to the separation in sky position with velocity,
the channel maps (not shown here) are correspondingly well fit.

\begin{figure}[tb]
\centering
\includegraphics[height=6.0in]{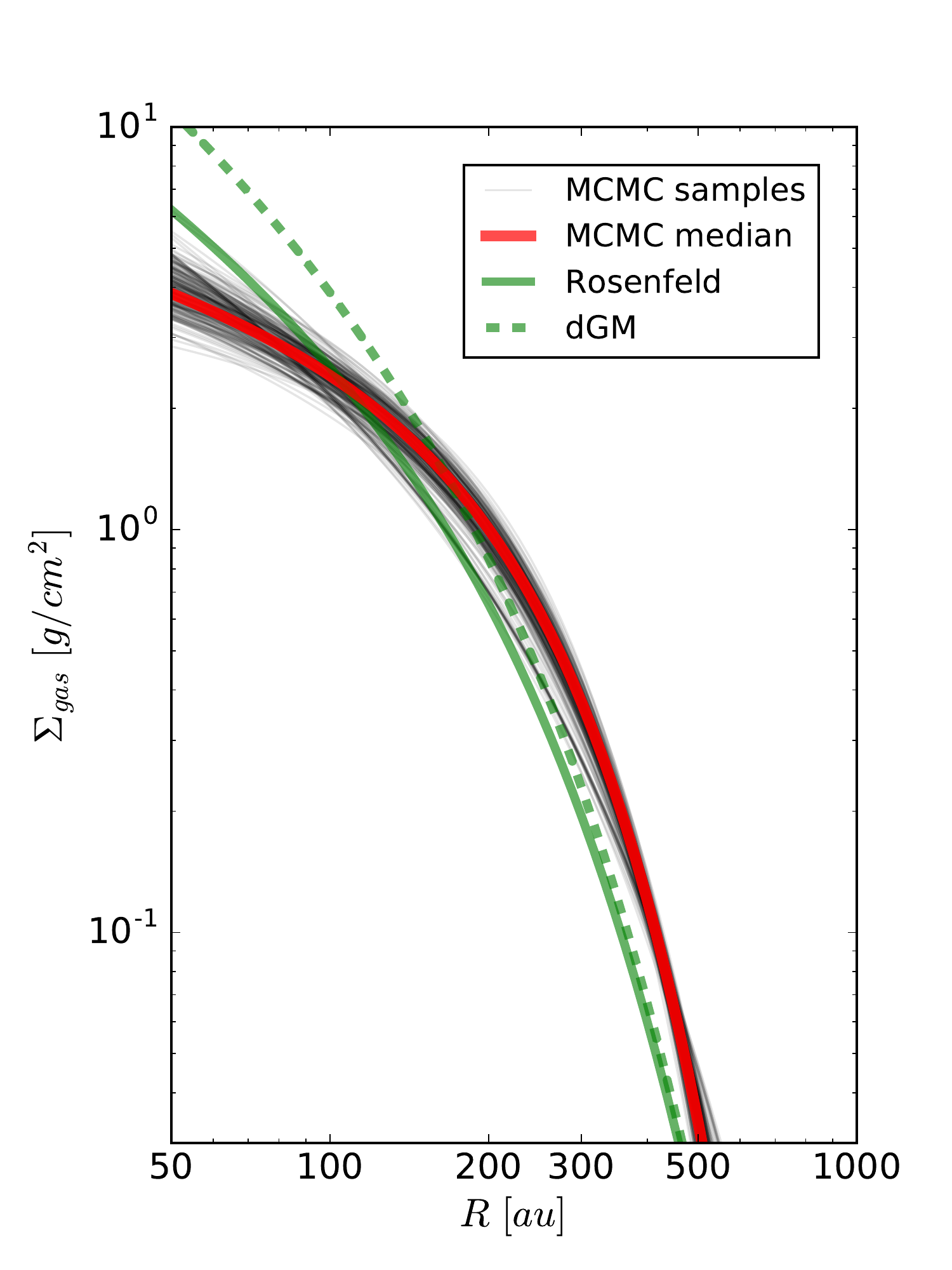}
\caption{
Gas surface density profiles derived from fitting the
the HD\,163296 \13CO\ 2--1 integrated map.
200 samples from the $5\times 10^4$ MCMC samples
are shown in gray to illustrate the range probed by
the random walk. The profile for the median
parameter values is shown in red.
For comparison, the profiles derived from fitting the
\twelveCO\ 3-2 channel maps by
\citep{2013ApJ...774...16R, 2013A&A...557A.133D}
are shown in green. The differences, which are due to the
higher optical depth of the \twelveCO\ line and the
nature of the fits, are discussed further in the text.
}
\label{fig:MCMC_profile}
\end{figure}

The inferred surface density density profiles, $\Sigma_{\rm gas}(R)$,
are shown in Figure~\ref{fig:MCMC_profile}.
For comparison, we also plot the surface density profiles
that were determined from fitting the \twelveCO\ 3-2 data by
\citet{2013ApJ...774...16R} (after normalizing to the same [CO]/[H$_2$]
abundance) and \citet{2013A&A...557A.133D}.
Both find higher central densities, steeper profiles,
and smaller outer disk radii than our fits to \13CO.
This cannot be due to higher optical depth in the
\twelveCO\ line nor to selective photo-dissociation of \13CO\
in the outer parts of the disk, both of which would produce the
opposite effect seen here, i.e., a larger CO and smaller \13CO\ disk.
More likely, it simply reflects the uncertainties inherent in
modeling a single line.

\subsection{Comparison with the C$^{18}$O image}
\begin{figure*}[tb]
\centering
\includegraphics[width=6.5in]{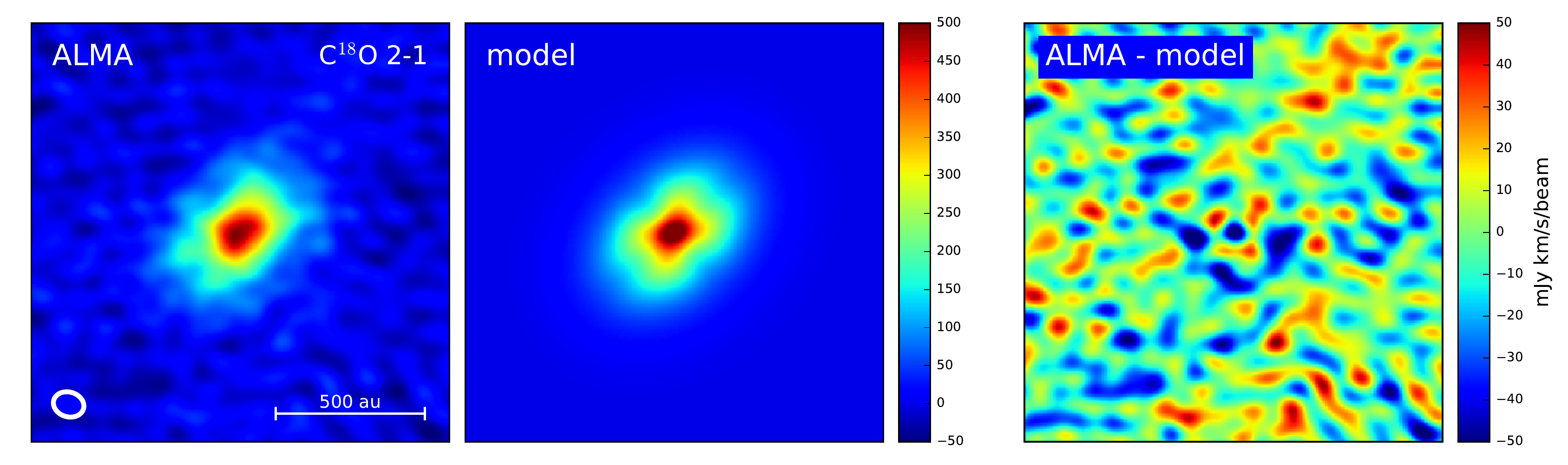}
\caption{
Comparison of \C18O\ integrated intensity images.
The left panel shows the ALMA Science Verification map,
the central panel the model fit for the disk parameter values
derived from the \13CO\ fitting,
and the right panel shows the difference image.
The colorbars show the range of intensities for each map;
the ALMA and model images on the same scale from
$-50$ to 500\,mJy\,\kms\,beam\e,
and the difference image from $-50$ to 50\,mJy\,\kms\,beam\e.
}
\label{fig:difference_image18}
\end{figure*}

The ALMA Science Verification data for HD\,163296
also include the \C18O\ 2--1 line
which, being of lower optical depth than the same \13CO\ transition,
allows a further test of the model.
We ran the RADMC-3D radiative transfer for this \C18O\ line 
with the median surface density parameters derived above and
only varied the [CO]/[\C18O] abundance to minimize the least
squares difference with the zero-moment map.
The data, image, and difference in Figure~\ref{fig:difference_image18}.
The fit is very good with peak residuals at about 10\% of the image
maximum. The inferred abundance ratio is [CO]/[\C18O]=700 and
there are no obvious systematics in the difference image suggesting
that the abundance does not greatly vary with disk radius.

As a rare isotopologue, \C18O\ cannot self-shield as effectively
as CO and is expected to be selectively photo-dissociated
\citep{1988ApJ...334..771V}.
Indeed the comparison of \13CO\ and \C18O\ line luminosities in
Paper I showed evidence for this in Taurus disks.
Detailed thermo-chemical models confirm that this can have a
significant impact on the \C18O\ emission from low-mass disks though
not for such massive disks as HD\,163296 \citep{2014A&A...572A..96M}.
Future observations of T Tauri disks can examine this important
effect and assess whether it may explain the variation of oxygen
isotopes in the Solar System \citep{2011Sci...332.1528M}.

\section{T Tauri model grid}
\label{sec:TTdisk_grid}
\subsection{Description}
Most stars are lower mass than HD\,163296 and their dust disks
tend to be considerably smaller in both mass and size
\citep{2010ApJ...723.1241A}.
Paper I and \citet{2016arXiv160405719A} show that the median Class II
disk gas mass in Taurus and Lupus is small,
$\sim 1\,M_{\rm Jup} = 10^{-3}\,M_\odot$,
and that Minimum Mass Solar Nebula (MMSN) disks with masses
$\sim 10^{-2}\,M_\odot$ are rare.
Whereas most studies of disk structure to date have
naturally tended toward observations of bright (i.e., massive)
and large disks, recent work shows that some low-mass disks may
be very small, at least as measured in the continuum
\citep{2014A&A...564A..95P}.
Nevertheless, the ALMA Science Verification data of HD\,163926
analyzed in \S\ref{sec:HD163296} had, by today's standards,
low resolution and high noise.
If we scale by the dynamic range in spatial and intensity scales,
$\sim 10$ and 40 respectively,
it seems feasible that current and future ALMA observations should
be able to map the intensity profile of these lower mass disks
with sufficient fidelity to derive their gas surface densities.

To be more quantitative and assess the best combination of resolution
and noise level to measure gas profiles, we defined a generic model
grid based on the parameters of disks around T Tauri stars.
As with HD\,163296, we fix the stellar mass and temperature
structure under the expectation that they can be determined through
fitting \twelveCO\ observations.
Following the nomenclature in Paper I, their values are set to 
\begin{equation}
M_{\rm star} = 0.5\,M_\odot,
~T_{\rm mid,1} = 100\,{\rm K},
~T_{\rm atm,1} = 500\,{\rm K},
~q = 0.5.
\end{equation}
\twelveCO\ kinematics, or even the continuum image, will also provide
a good measure of the disk inclination to the line of sight.
As the inclination changes the image surface brightness,
however, it affects our ability to measure gas surface
density profiles and we therefore consider a range of values
in the grid. The set of surface density parameters
and inclination are listed in Table~\ref{tab:TTauri_grid}.
For comparison with planned and ongoing ALMA surveys,
we created zero-moment maps of the \13CO\ 2--1 line
for a distance of 150\,pc\ and a pixel scale of 5\,au ($0\farcs 03$).

\begin{deluxetable}{cccc}
\tabletypesize{\footnotesize}
\centering
\tablecolumns{3}
\tablewidth{0pt}
\tablecaption{Parameter values of the T Tauri grid \label{tab:TTauri_grid}}
\tablehead{
\colhead{Parameter} & \colhead{Range} & \colhead{Step} & \colhead{Units}
}
\startdata
$M_{\rm gas}$       & 0.1--3.0 &   0.1   & $10^{-2}\,M_\odot$ \\
$R_c$               & 20--200  &  20     & au                 \\
$\gamma$            & 0.0--1.0 &   0.1   &                    \\
inclination         & 30--60   &  15     & $^\circ$           
\enddata
\end{deluxetable}

\begin{figure}[tb]
\centering
\includegraphics[width=6.5in]{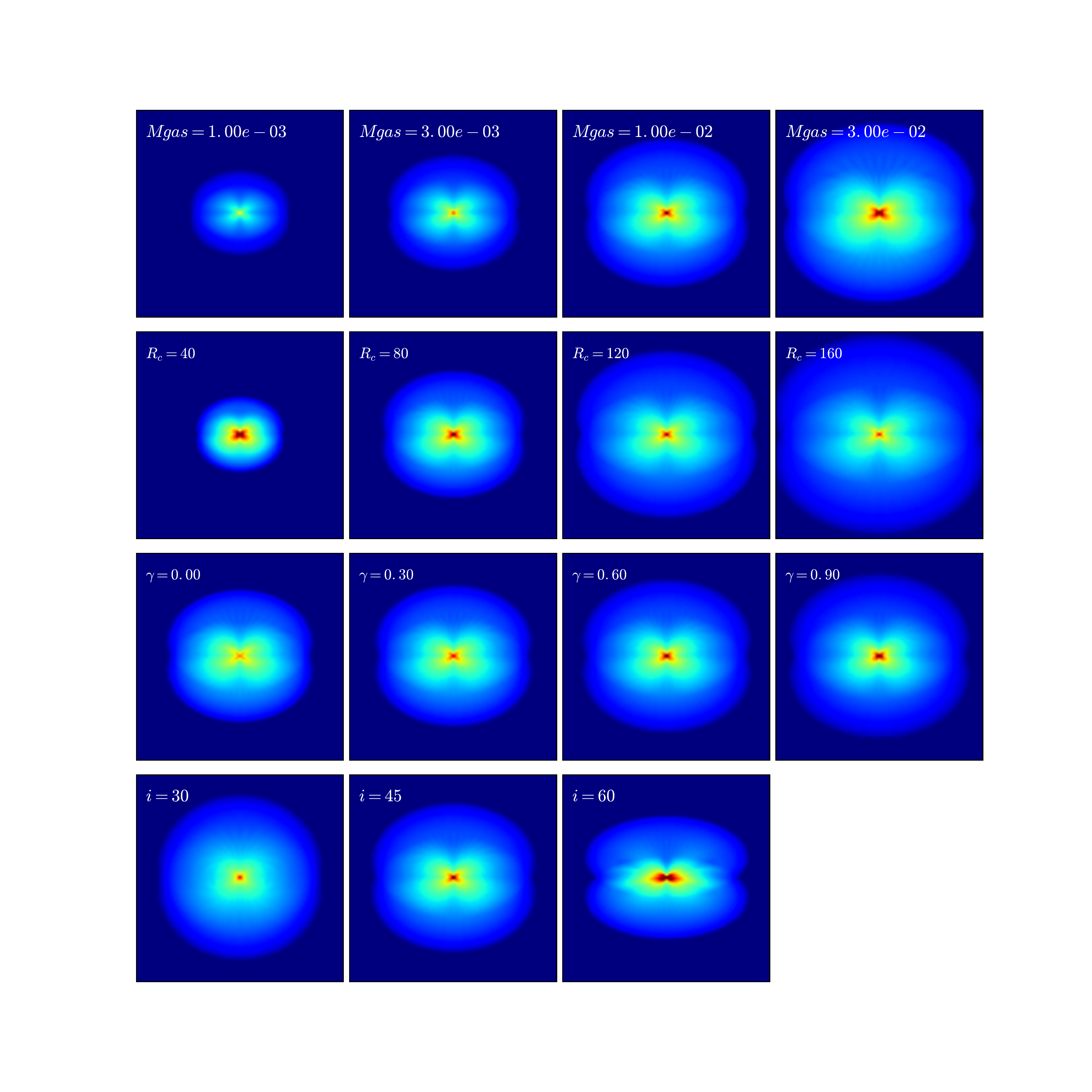}
\caption{
Montage of \13CO\ 2--1 integrated intensity maps from the
T Tauri disk model library.
Each row shows the effect of varying a single parameter,
whose value is shown in the upper left of each subplot.
Unless otherwise labeled, the default values are
$M_{\rm gas}=10^{-2}\,M_\odot, R_c=100\,{\rm au},
\gamma=0.5, i=45^\circ$.
The images are shown with square-root scaling varying from
0 to 100\,mJy per 5\,AU$\times$5\,AU cell size.
}
\label{fig:TTdisk_montage}
\end{figure}

The resulting image library consists of 9900 models.
Figure~\ref{fig:TTdisk_montage} shows the effect of
varying the three surface density parameters and inclination.
$M_{\rm gas}$ scales the intensity at each pixel, though not
perfectly linearly or uniformly across an image due to opacity.
Disks with small $R_c$ are compact with a high central brightness
whereas the largest disks have low surface brightness.
The center also brightens as the density gradient $\gamma$ increases.
The bottom row shows the strong effect of inclination on
the disk aspect ratio and brightness distribution.
We assume this to be fixed in deriving the three other
parameters but allow for its variation when considering
the statistical accuracy of a survey.

\subsection{Criteria for distinguishing a disk from a gaussian}
The gas surface density parameters each have a different effect
on the \13CO\ zero-moment map and each can, in principle,
be distinguished from one another.
As observations are inherently noisy and of finite resolution, however,
the determination of parameters will not be exact and may be biased.
We can assess the effects of parameter variation, at least within the
limitations of our model framework, by convolving the images and
adding gaussian random noise.

ALMA can now readily achieve a resolution from $\ll 0\farcs 1$ to
about an arcsecond at the 220\,GHz frequency of \13CO\ 2--1.
Independent of the array configuration, the noise per beam
for a given frequency depends only on the integration time
but as the flux per beam decreases with increasing resolution
for a resolved source, there is a tradeoff between spatial and
intensity dynamic range.
Based on the model disk sizes and line fluxes, we consider a range
of (circular) beamsizes from $0\farcs1$ to $0\farcs4$ FWHM,
and rms noise levels from 1 to 30\,mJy\;beam\e\;\kms.
As a guide to the array integration time, the ALMA sensitivity
calculator\footnote{https://almascience.nrao.edu/proposing/sensitivity-calculator},
shows that an rms of $\sigma = 10$\,mJy\;\kms\;beam\e\
at 220\,GHz can be achieved in 2 minute integrations at
$-30^\circ$ declination.

Before comparing a simulated image against the model library,
it is important to test the ansatz that the image can be
distinguished as a disk and not a simpler gaussian description.
We fit an elliptical gaussian to a simulated image\footnote{We fix the
offsets of fit to (0,0) as we expect the disk center to be well defined
from the high signal-to-noise continuum image.}
and calculate the flux distribution of the residuals.
A Kolomogorov-Smirnov test of the goodness of fit of a
gaussian to the residual flux distribution then shows how
well the image can be distinguished from the elliptical
gaussian fit.

\begin{figure}[tb]
\centering
\includegraphics[width=5.0in]{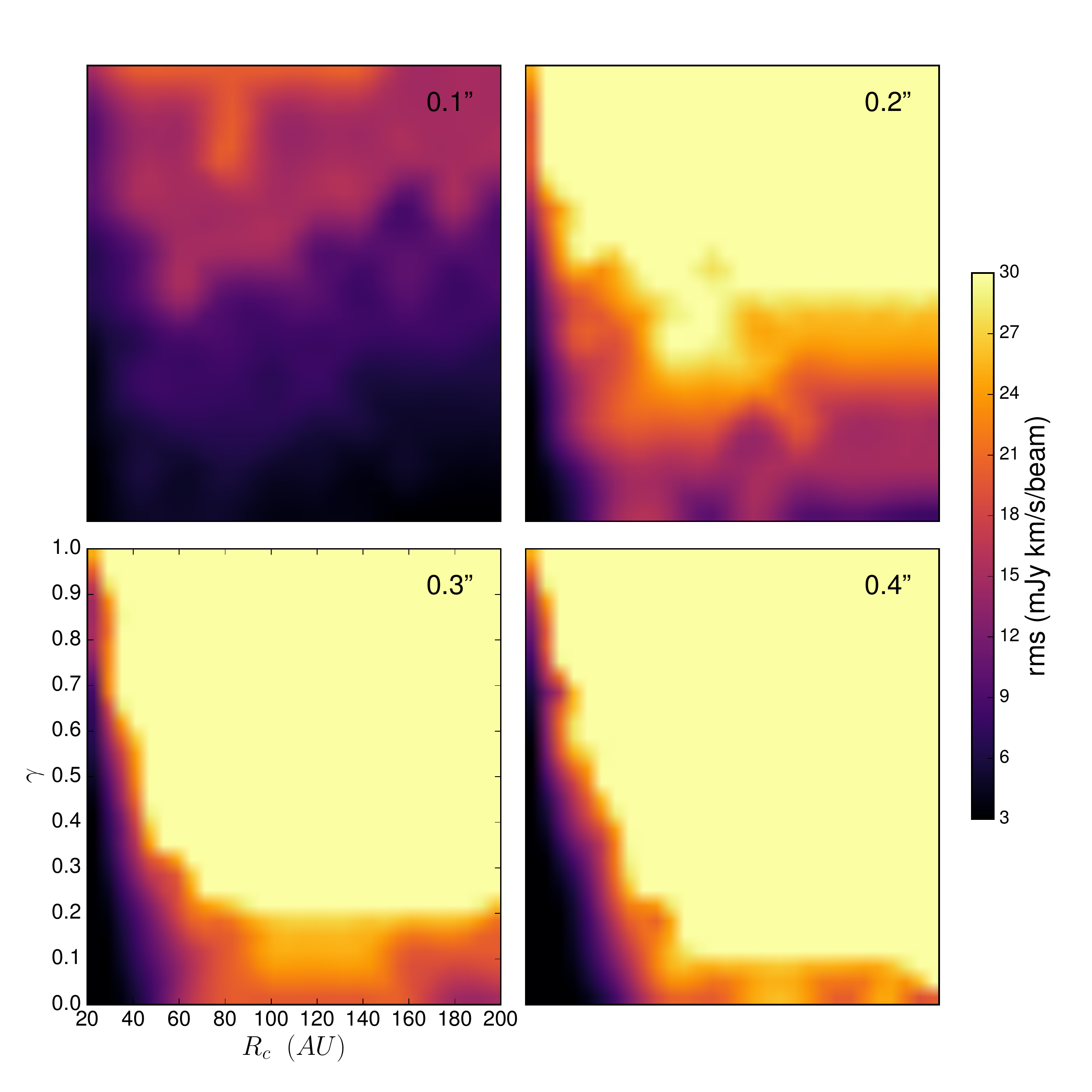}
\caption{
Plots of the rms noise level required to distinguish a
T Tauri \13CO\ 2--1 disk model with $M_{\rm gas}=10^{-2}\,M_\odot$
from an elliptical gaussian fit. Each plot is shown as
a function of $R_c$ and $\gamma$ and the different panels
have different beam sizes, as labeled in the upper right corner.
Large disks and steep profiles are readily distinguished from
a gaussian at moderate resolution, $0\farcs 3-4$,
but small and/or flatter disks require higher resolution and low rms.
}
\label{fig:gauss_MMSN}
\end{figure}

For each convolved disk model, we increase the rms noise
until a gaussian fit is sufficient. This noise level is plotted
for the four different beam sizes and MMSN disks
($M_{\rm gas}=10^{-2}\,M_\odot$)
as a function of $R_c$ and $\gamma$ in Figure~\ref{fig:gauss_MMSN}.
The bright yellow regions show where disks are readily
distinguished as disk-like even in shallow integrations with
relatively high noise levels, 30\,mJy\;\kms\;beam\e.
There is a balance between resolution and signal-to-noise.
A smaller beamsize provides more independent measurements of
the disk shape and is necessary to resolve compact disks,
but there is less flux per beam.
The smallest beamsize shown here, $0\farcs 1$, is so fine that
very sensitive observations are necessary to image the gas structure.
For larger beamsizes, the purple regions indicate that an
rms of 10\,mJy\;\kms\;beam\e\ can distinguish all but the
smallest disks, $R_c\lesssim 40$\,au, with flat profiles, $\gamma\lesssim 0.5$.
Not surprisingly, this unresolved region is larger for the
largest beam size, $0\farcs 4$.

\begin{figure}[tb]
\centering
\includegraphics[width=5.0in]{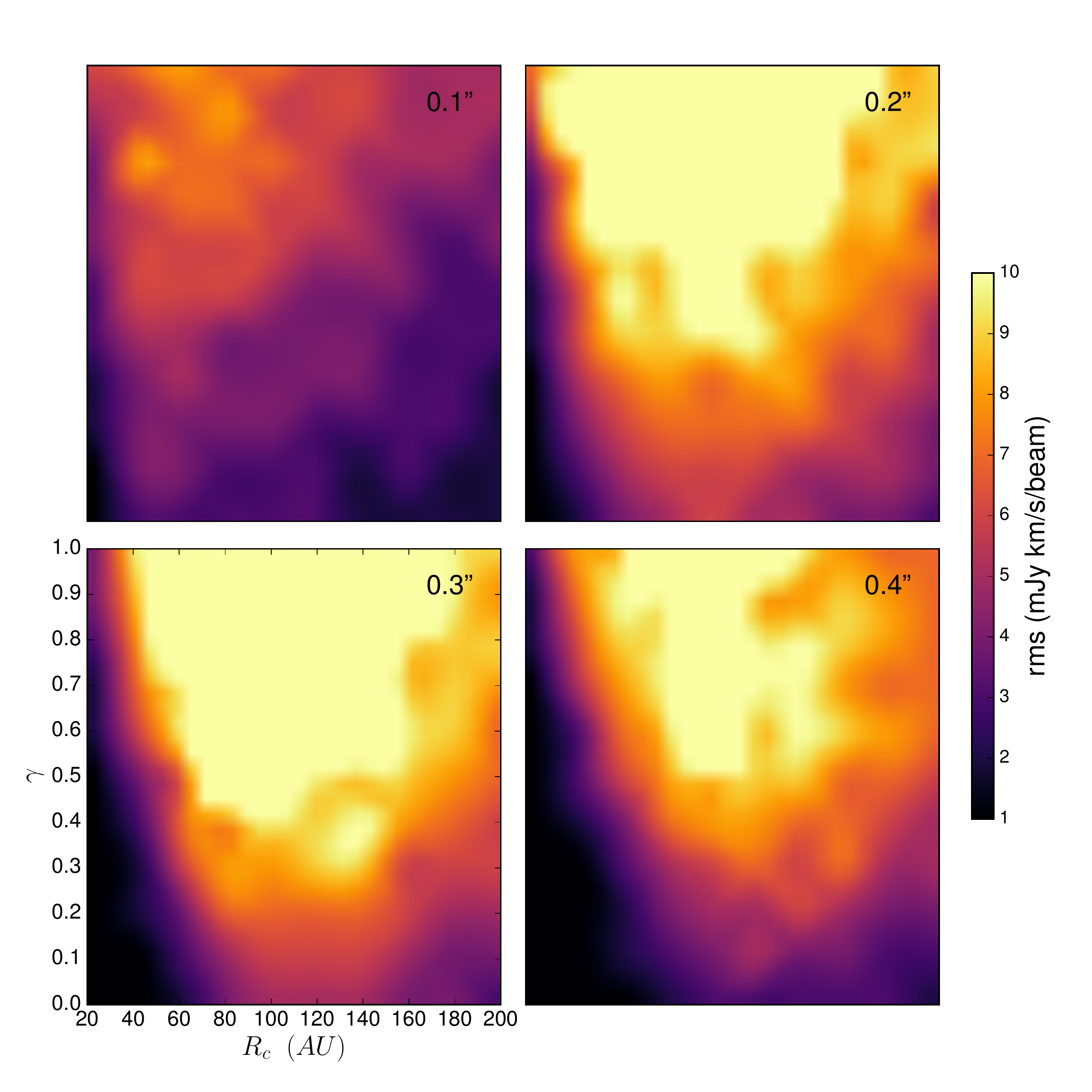}
\caption{
Similar plot as Figure~\ref{fig:gauss_MMSN}
but for a Jupiter mass disk, $M_{\rm gas}=10^{-3}\,M_\odot$.
Note the change in color scale for the rms.
Due to the much lower flux levels, large disks are hard
to distinguish from a gaussian even at the largest beam
size here, $0\farcs 4$, and the optimum range is an
intermediate resolution, $0\farcs 2-3$.
Even then, low noise levels are required to study
these low mass disks.
}
\label{fig:gauss_MJup}
\end{figure}

The fainter, lower mass disks require more sensitive observations.
Figure~\ref{fig:gauss_MJup} shows the same calculation as above
for but a Jupiter mass disk, $M_{\rm gas}=10^{-3}\,M_\odot$,
and a color scale scaled lower by a factor of three.
As before, the signal-to-noise level in a $0\farcs 1$ beam is too
low to study the gas structure, and small/flat disks are
inaccessible to all but the most sensitive observations.
However, unlike the MMSN disks, large and flat disks
$R_c\gtrsim 160$\,au, $\gamma\lesssim 0.3$ are also hard
to study due to their low, extended surface brightness.
For these low mass disks, an intermediate sized beam,
$0\farcs 2-0\farcs 3$ and low noise levels,
$\sim 3$\,mJy\;\kms\;beam\e\ ($\sim 20$\,minute integrations)
are optimal.

\subsection{Accuracy of parameter estimation}
If we can observe a disk with sufficient resolution and sensitivity
to differentiate it from a gaussian, the next question is how well
can we determine the gas surface density profile.
We use the same image interpolation scheme described in
\S\ref{sec:image_interpolation} to continuously sample the
parameter space and find the best fit to a given simulated image
using the python {\tt scipy.optimize} routine.
We do not estimate errors in this case but rely on the statistics
of the model comparisons to assess the accuracy to which we can
measure each parameter.

\begin{figure}[tb]
\centering
\includegraphics[height=5.5in]{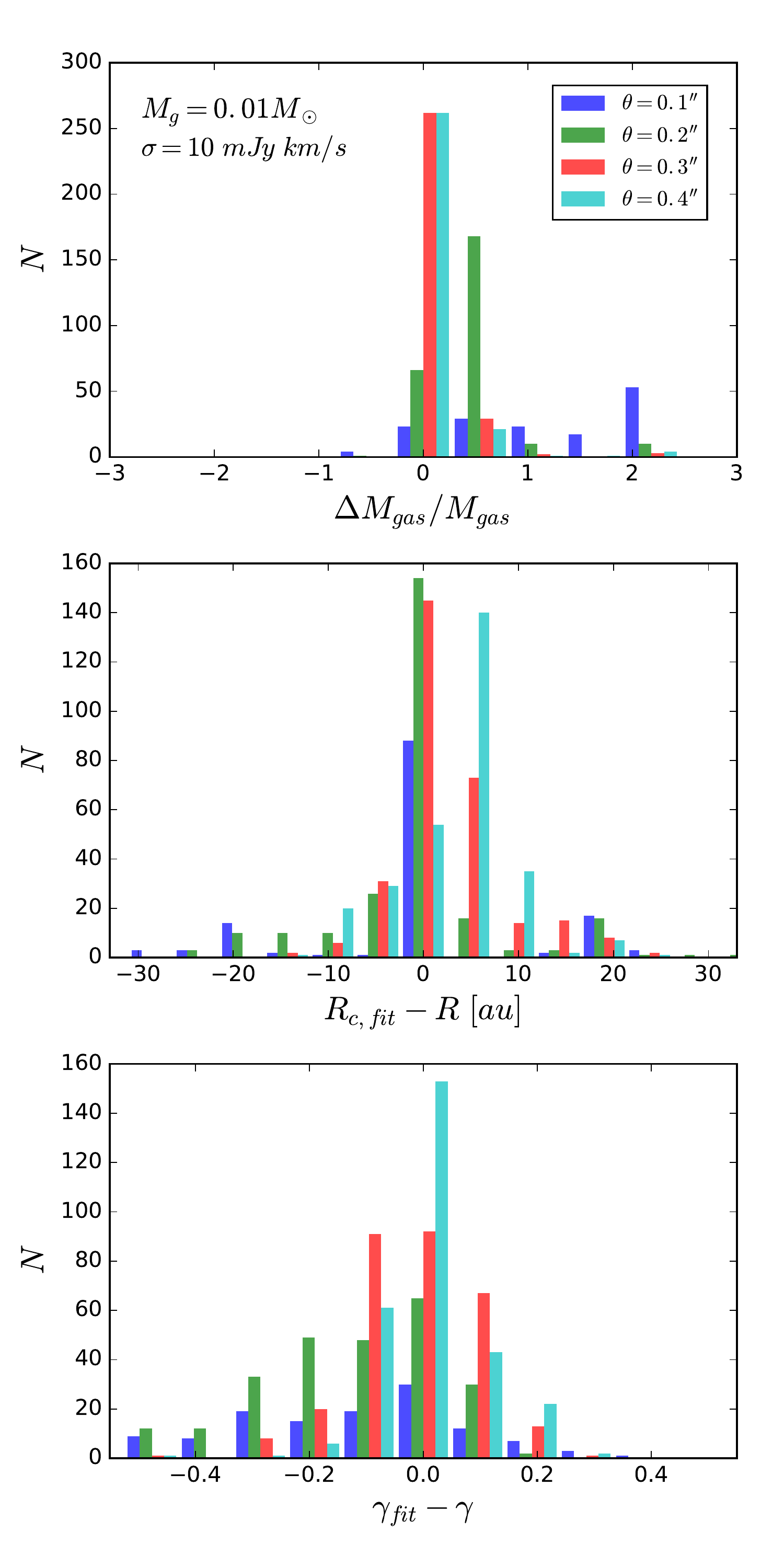}
\caption{
The precision with which the gas surface density profile of a disk
can be measured, color-coded by resolution. The model disks have a
gas mass, $M_{\rm gas}=10^{-2}\,M_\odot$, and the noise in the
simulated observations is 10\,mJy\,\kms\,beam\e.
The top panel shows the fractional precision to which we
recover the input gas mass.
The two lower panels show the absolute
difference between the input and best fit $R_c$ and $\gamma$.
As only those disks
which can be distinguished from a gaussian are shown, there are
fewer models at the $\theta_{\rm FWHM} = 0\farcs 1$ beam size
(see Figure~\ref{fig:gauss_MMSN}).
}
\label{fig:TTdisk_fit_MMSN}
\end{figure}

Figure~\ref{fig:TTdisk_fit_MMSN} plots histograms of the
difference between the input and fitted parameters
for all disks with mass $M_{\rm gas}=10^{-2}\,M_\odot$,
observed with an rms noise level of 10\,mJy\;\kms\;beam\e.
The histograms are color-coded by beam size and only
those disks that can be distinguished from a gaussian
are shown. Hence there are fewer disks in the dark blue
($\theta_{\rm FWHM}=0\farcs 1$).
The top panel shows the relative difference between the
fitted gas mass and the input value. In general the
mass is measured from the profile fitting alone to within
about 50\% for all but the noisy, high resolution images.
The two lower panels show the absolute difference between the
input and measured $R_c$ and $\gamma$.
The characteristic radius is generally measured to less than
10\,au at all beam sizes, with the best results for a
$0\farcs 2$ beam, and a slight bias toward overestimating
sizes at the lowest resolution here, $0\farcs 4$.
The gradient, $\gamma$, is most accurately measured at
$0\farcs 3$ (red histogram) which typically provides an
ideal combination of multiple resolution elements with high
signal-to-noise across the disk.

\begin{figure}[tb]
\centering
\includegraphics[height=5.5in]{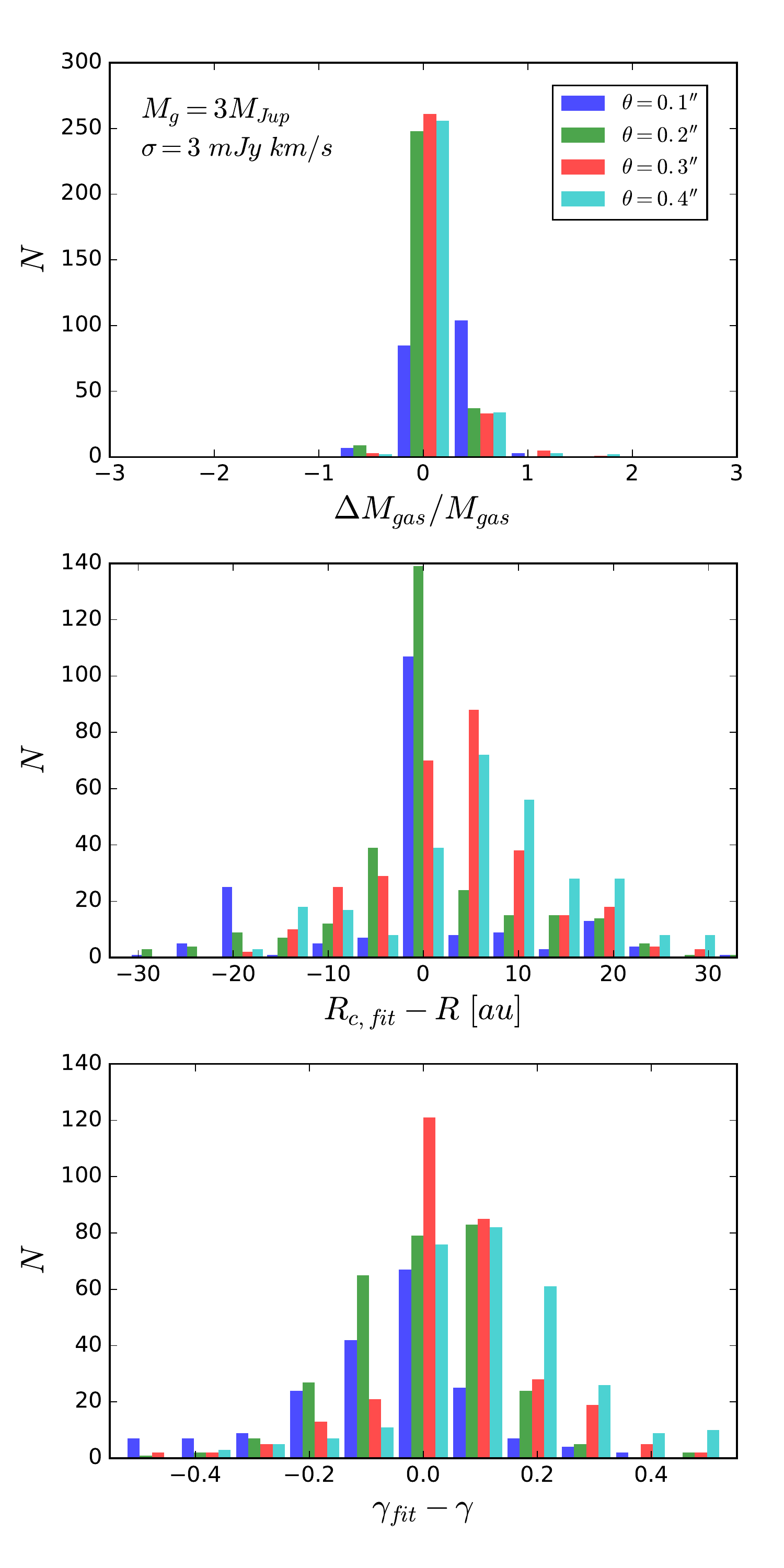}
\caption{
As Figure~\ref{fig:TTdisk_fit_MMSN} but for disks with a gas mass
$M_{\rm gas}=3\times 10^{-3}\,M_\odot$ and noise 3\,mJy\,\kms\,beam\e.
}
\label{fig:TTdisk_fit_3MJup}
\end{figure}

Figure~\ref{fig:TTdisk_fit_3MJup} plots histograms of
the fits for all the disks with a lower mass,
$M_{\rm gas}=3\,M_{\rm Jup} = 3\times 10^{-2}\,M_\odot$,
observed at a lower noise level, 3\,mJy\;\kms\;beam\e.
The results are similar to Figure~\ref{fig:TTdisk_fit_MMSN}
as might be expected given that the mass and noise level
decreased by the same factor of $\sim 3$.

In summary and as a general guideline, the best results are obtained
for an intermediate resolution, $0\farcs 2 - 0\farcs 3$,
and once a disk is observed with sufficient signal-to-noise
to distinguish it from a gaussian, we find that we can measure the
disk size and gradient parameters to within about
$\Delta R_c = 10$\,au, $\Delta\gamma = 0.2$.

\section{Discussion}
\label{sec:discussion}
Unlike the turbulent interstellar medium, the gas structure and kinematics
in protoplanetary disks is relatively simple and prescriptive.
The complexity in measuring gas masses and surface density profiles
resides in the chemistry and radiative transfer required to interpret
the observations. CO is an abundant, stable, and readily observable species.
Its formation uses almost all available gas-phase C and O and its destruction
follows two main pathways, photo-dissociation and freeze-out, that are
amenable to semi-analytical models. The intricacies of isotopologue
selective dissociation for \13CO\ is largely compensated by the
ion-molecule exchange reaction,
\twelveCO\ + $^{13}{\rm C}^+ \longrightarrow\ ^{12}{\rm C}^+$ + \13CO,
deep in the warm molecular layer \citep{2009A&A...503..323V}.
With a good balance between low optical depth and detectability,
\13CO\ is the molecule of choice for measuring the gas mass distribution.
Finally, because the ordered Keplerian rotation largely separates the
emission from different parts of a disk into different spectral channels,
we can compare model images to integrated intensity line maps without great
loss of information, simplifying and speeding up the fitting process
considerably.

Any mass or column density measurement that is derived from observations
of a trace molecule fundamentally relies on knowledge of that
molecule's abundance relative to H$_2$.
We assume that the [\twelveCO]/[H$_2$] abundance is the same
($10^{-4}$) in disks as in molecular clouds and cores.
There are, unfortunately, few tests of this and they disagree.
\citet{2014ApJ...794..160F} directly measure the abundance
from absorption lines through a flared, inclined disk
and show agreement with the ISM value.
On the other hand, comparison of HD and CO isotopologue lines
in the TW Hya disk led \citet{2013ApJ...776L..38F} to a much
lower abundance. They attributed this to an active carbon chemistry
that removes CO from the warm molecular layer and locks up
volatiles on large dust grains in the cold midplane
\citep[see also][]{2016A&A...588A.108K}.
This is a fascinating suggestion that should be testable
with more complete inventories of disk gas and statistical
studies of gas evolution.
Of course, any uncertainty in the global CO abundance translates into
the normalization, but not the shape, of the surface density profile.

Our ability to fit the integrated intensity map of \13CO\ 2--1
in the large, bright HD\,163296 disk demonstrates the feasibility
of our modeling procedure. Although our formal errors were small,
our derived surface density profile differs from fits to the
\twelveCO\ 3--2 map by
\citet{2013ApJ...774...16R} and \citet{2013A&A...557A.133D}.
The \twelveCO\ line has a much higher optical depth, however,
and and the primary focus of these two studies was on the
temperature rather than density structure.
A more holistic approach would be to analyze both lines to
simultaneously determine the temperature and density structure.
By allowing for variation in the the temperature,
we would also expect larger errors in the surface density parameters
than we report in this proof-of-concept study.

Most stars are lower mass than HD\,163926 and most disks are
corresponding less massive and also smaller.
The first large study designed to measure disk gas masses in
a representative sample is the ALMA Lupus survey by
\citet{2016arXiv160405719A}.
They found a very low median gas mass,
$M_{\rm gas} \sim 1\,M_{\rm Jup} = 10^{-3}\,M_\odot$,
and the \13CO\ maps have much lower image fidelity than
that of HD\,163296.
The T Tauri grid described in \S\ref{sec:TTdisk_grid}
shows that we can extend the same modeling technique,
at least for the upper end of that sample,
$M_{\rm gas} \gtrsim 3\,M_{\rm Jup}$,
through higher resolution, higher sensitivity observations.
For the \13CO\ 2--1 line, the requirement is
a resolution of $0\farcs 2 - 0\farcs 3$
and a (mass dependent) noise level of 3--10\,mJy\;\kms\;beam\e.
The lower rms is achieved with ALMA in 20 minutes
so line imaging surveys to determine disk gas surface density
profiles are quite feasible in moderate amounts of time.
Furthermore the 2--1 lines of both \twelveCO\ and \C18O\
can be simultaneously observed with the Band 6 receivers.
This is potentially a very powerful combination that permits
the co-modeling of temperature and density and the study
of selective photo-dissociation. The determination of the
gas properties in this way would also provide an essential
reference for measuring the abundance and distribution of 
other molecules.

There have been numerous surveys of the continuum emission from
disks and analyses of their solid content.
Studies of the disk gas content and distribution provides an
additional observational dimension for following their diverse
evolutionary pathways.
As we gain a more complete picture of both components, gas and dust,
we can hope to better understand planet formation and the tremendous
range of exoplanet types.

\acknowledgments
This paper makes use of the following ALMA data:
ADS/JAO.ALMA\#2011.0.00010.SV.
ALMA is a partnership of ESO (representing its member states),
NSF (USA) and NINS (Japan), together with NRC (Canada),
NSC and ASIAA (Taiwan), and KASI (Republic of Korea),
in cooperation with the Republic of Chile.
The Joint ALMA Observatory is operated by ESO, AUI/NRAO and NAOJ.
JPW is supported by funding from the NSF and NASA
through grants AST-1208911 and NNX15AC92G.
We thank the community developers of the Python packages for Astronomy
in Astropy \citep{2013A&A...558A..33A}.

{\it Facilities:} \facility{ALMA}.


\end{document}